\documentclass[aps,twocolumn,amsmath,amssymb,preprintnumbers,floatfix,prb,superscriptaddress,longbibliography]{revtex4-2}

\usepackage{comment}
\usepackage[version=4]{mhchem}
\usepackage[utf8]{inputenc}
\usepackage{newtxtext}
\usepackage[upint]{newtxmath}
\usepackage{microtype}
\usepackage{textcomp}
\usepackage{dsfont}
\usepackage{eucal}
\usepackage{siunitx}
\usepackage{soul}

\usepackage{enumerate}
\usepackage{amsfonts}
\usepackage{color}
\usepackage{soul}

\usepackage{todonotes}
\presetkeys%
    {todonotes}%
    {inline}{}

\usepackage{graphicx}

\usepackage[colorlinks,allcolors=blue]{hyperref}
\usepackage[capitalize]{cleveref} 
\usepackage{cleveref}

\newcommand{\vecR}{\boldsymbol{R}}

\renewcommand{\vec}[1]{\boldsymbol{#1}}

\newcommand{\vecr}{\boldsymbol{r}}      

\definecolor{DarkBlue}{rgb}{0,0,0.80}
\definecolor{DarkRed}{rgb}{0.80,0,0}
\definecolor{Purple}{rgb}{0.55,0,0.55}

\newcommand{\jacob}[1]{\textcolor{Purple}{#1}}

\begin{document}

\title{Quantum interference between vortex- and impurity-bound states boosts thermoelectricity}

\author{Erik Haatuft}
\affiliation{Center for Quantum Spintronics, Department of Physics, Norwegian \\ University of Science and Technology, NO-7491 Trondheim, Norway}
\author{Jacob Linder}
\email{jacob.linder@ntnu.no}
\affiliation{Center for Quantum Spintronics, Department of Physics, Norwegian \\ University of Science and Technology, NO-7491 Trondheim, Norway}

\begin{abstract}
Thermoelectric effects in superconductors are generally suppressed by the approximate particle-hole symmetry of the quasiparticle spectrum, but can become pronounced near defects that host particle-hole asymmetric bound states. Here, we investigate how the local thermoelectric response is modified when multiple vortices or nonmagnetic impurities are brought into proximity. Using a lattice Bogolioubov-de Gennes approach combined with linear-response tunneling theory, we calculate the spatially resolved density of states and Seebeck coefficient in $s$- and $d$-wave superconductors. We find that the thermoelectric response can be strongly enhanced when spatially extended defect-induced states overlap. For vortices, this enhancement persists beyond the immediate core regions and originates from interference between vortex-bound states. For impurities, the thermoelectric response exhibits a comparable dependence on impurity separation in the $s$- and $d$-wave cases, although the associated spectral reconstruction is considerably more localized in the $s$-wave superconductor. Our results show that the spatial extent and interference of defect-induced quasiparticle states provide a means of controlling local thermoelectricity in inhomogeneous superconductors, with potential relevance for cryogenic thermoelectric sensing and energy conversion.
\end{abstract}

\maketitle

\section{Introduction}
Thermoelectric phenomena convert temperature differences into electrical signals and vice versa, providing the operating principle of thermoelectric generators, coolers, and thermal sensors \cite{goldsmid_introduction_2016,qin_solid-state_2022}. Although superconductors are restricted to low operating temperatures, controlling their thermoelectric response is relevant for cryogenic energy conversion and sensing applications \cite{heikkila_pss_19}. In the linear-response regime, a finite thermoelectric effect requires particle-hole asymmetry in the electronic spectrum near the Fermi level \cite{marchegiani2020nonlinear}. The quasiparticle spectrum of a homogeneous BCS superconductor is approximately particle-hole symmetric, strongly suppressing its conventional thermoelectric response \cite{benenti_fundamental_2017}. Spatial inhomogeneities provide a way to overcome this restriction by generating local spectral asymmetry.

One important class of such inhomogeneities is formed by Abrikosov vortices in type-II superconductors \cite{abrikosov_magnetic_1957, tinkham_book}. The superconducting order parameter is suppressed in the vortex core and acquires a phase winding around it, giving rise to vortex-bound quasiparticle states \cite{caroli_bound_1964}. These states extend over a finite region surrounding the vortex core and can produce a locally particle-hole asymmetric density of states. This spectral asymmetry was recently predicted to generate a large local thermoelectric response \cite{singh_prl_2024}. Vortices, however, generally do not occur in isolation, but form configurations whose density and separation can be controlled by the applied magnetic field. When the spatially extended states associated with different vortex cores overlap, their interference can modify both the quasiparticle spectrum and its particle-hole asymmetry. It is therefore natural to ask whether the thermoelectric response of a multi-vortex system can be understood as a superposition of independent single-vortex contributions, or whether interference between vortex-bound states gives rise to a collective response.

Impurities provide a second source of spatial inhomogeneity, whose effect depends strongly on the superconducting pairing symmetry. In a conventional $s$-wave superconductor, a nonmagnetic impurity potential generally produces a strongly localized spectral reconstruction, typically near the superconducting gap edge \cite{machida_bound_1972, shiba_hartree-fock_1973, balatsky_impurity-induced_2006}. In a $d$-wave superconductor, by contrast, the sign-changing order parameter allows a nonmagnetic impurity to generate pronounced low-energy resonances \cite{balatsky_impurity-induced_2006}. The thermoelectric response of a $d$-wave superconductor containing a single impurity has previously been shown to depend strongly on the impurity potential, vanishing in the Born and unitary limits while becoming large in the intermediate-scattering regime \cite{lofwander_large_2004}. For several impurities, the spatial extent of the impurity-induced states determines whether the response remains localized to the individual defects or if it is modified by overlap and interference between them. A comparison between vortices and impurities can therefore reveal how the spatial structure of defect-induced quasiparticle states controls thermoelectricity in inhomogeneous superconductors.

\begin{figure*}[t]
    \centering
    \includegraphics[width=\linewidth]{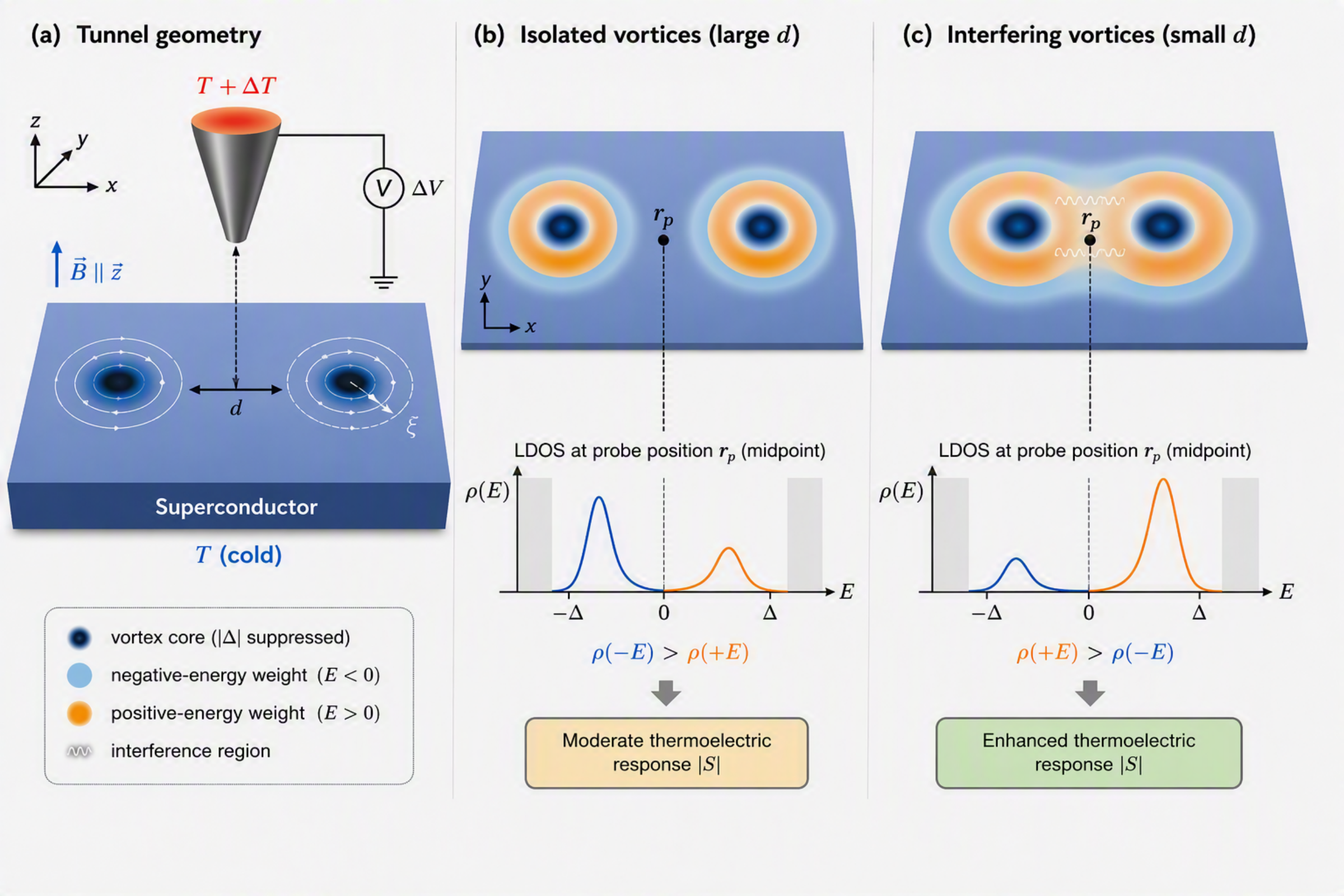}
    \caption{Qualitative illustration of the system and the mechanism underlying the enhanced thermoelectric response. (a) A temperature-biased normal-metal probe couples locally to a type-II superconductor containing two vortices separated by a distance $d$. The thermally induced voltage difference $\Delta V$ determines the local Seebeck response. (b) For well-separated vortices, the spatially extended vortex-bound states are approximately independent. (c) At smaller separation, the vortex-bound states overlap and interfere, reconstructing the local spectrum and enhancing its particle-hole asymmetry. In (b) and (c), the tilted surfaces represent the superconducting $xy$-plane, while the orange and blue regions schematically indicate local dominance of positive- and negative-energy spectral weight, respectively. The LDOS curves are evaluated at the same relative probe position $\mathbf r_{\mathrm p}$, chosen as the midpoint between the vortices. The difference between the two LDOS curves therefore results from the reduced vortex separation rather than from a change in probe position.}
    \label{fig:model}
\end{figure*}

\begin{figure*}[t]
    \centering
    \includegraphics[]{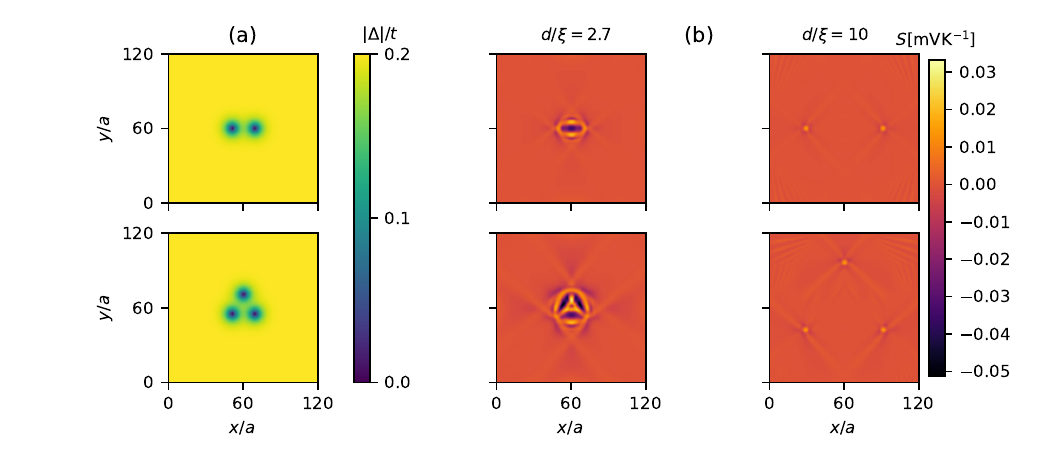}
    \caption{(a): The chosen superconducting OP profiles, $|\Delta(\vecr)|$, for two (top) and three (bottom) vortices. We have chosen the bulk OP $\Delta_0=0.2t$ and the superconducting coherence length $\xi_0=6a$. (b): Two-dimensional heatmap of the spatially resolved Seebeck coefficient in an $s$-wave superconductor. Here we consider two cases: (left) the vortices are placed in close proximity at their minimum separation distance $d_{\text{min}}$, and (right) when they are placed far apart.}
    \label{fig:OP_Seebeck_2D}
\end{figure*}

In this work, we investigate how the local thermoelectric response of $s$- and $d$-wave superconductors is modified by the presence of multiple vortices or nonmagnetic impurities. Using a lattice Bogoliubov--de Gennes description combined with linear-response tunneling theory, we calculate the spatially resolved density of states and Seebeck coefficient for different defect configurations and separations. We find that bringing the defects closer together can strongly enhance the thermoelectric response through overlap and interference between their bound states, which reconstruct the local spectrum and increase its particle-hole asymmetry. In the vortex case, this effect extends beyond the immediate core regions because of the spatially extended nature of the vortex-bound states. The impurity calculations demonstrate that the spatial structure of the response depends on the pairing symmetry, with a more localized modification of the order parameter and spectral properties in the $s$-wave case than in the $d$-wave case. Our results identify defect separation and the spatial extent of defect-induced quasiparticle states as key factors controlling local thermoelectricity in inhomogeneous superconductors. For vortices, the comparison with a minimal hybridization model further identifies bound-state interference as the microscopic origin of the enhanced response.

\section{Methodology} 
We focus here on the Seebeck coefficient $S = -\Delta V/\Delta T$, where $\Delta V$ is the generated electric voltage difference due to an applied thermal bias $\Delta T$ in an open-circuit setup. This ratio can be computed using linear response theory as $S = \frac{L_{12}}{T L_{11}}$ where the Onsager transport coefficients are given by \cite{singh_prl_2024} 
\begin{align}
    L_{11} &= -gT \int dE \sum_i |t(\vecr_i)|^2 \rho (\vecr_i, E) f'(E),  \label{eq:L11} \\
    L_{12} &= -\frac{gT}{e} \int dE \sum_i |t(\vecr_i)|^2 \rho(\vecr_i, E) f'(E) E, \label{eq:L12}
\end{align}
where $g$ is a constant and $f'(E)$ is the derivative of the Fermi-Dirac distribution. Here, quantum tunneling occurs between a voltage-biased STM-tip and the bound states where $t(\vecr)$ describes the spatial profile of the tip relative to the superconducting surface. Microscopically, this factor originates from the tunneling matrix element between states in the metallic tip and superconductor. To model an STM tip, we choose a Gaussian profile centered at the tip position $\vecr_p$
\begin{equation}
    |t(\vecr)|^2 = A \ e^{-|\vecr - \vecr_p|^2/(2R_{\text{STM}}^2)},
\end{equation}
where we set the effective tip radius $R_{\text{STM}}=a$, where $a$ is the lattice constant. Note that the prefactor $A$ is unimportant in our calculations, as it cancels in the ratio $L_{12}/L_{11}$.
To compute the Onsager coefficients, we need to determine the spectral electronic properties of the system by obtaining the local density of states, $\rho(\vecr, E)$. This can be done in a fully quantum mechanical way through the lattice Bogolioubov-de Gennes framework.

\begin{figure*}[t]
    \centering
    \includegraphics[]{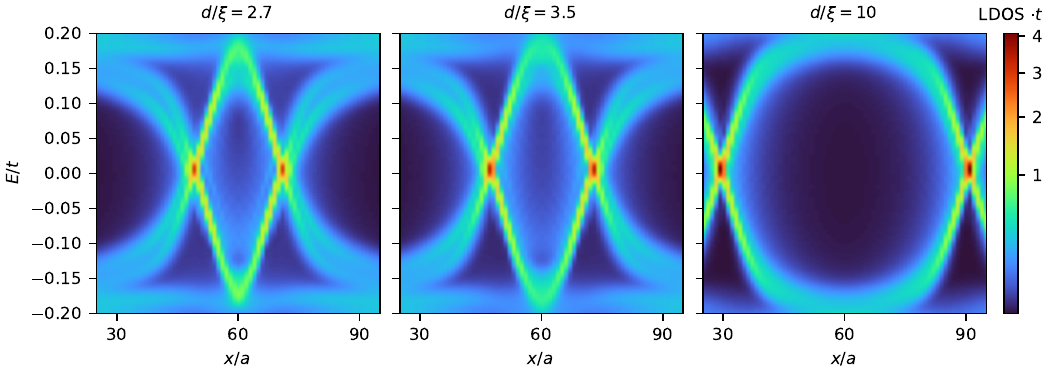}
    \caption{LDOS as a function of energy and $x$ along the intervortex line. The vortices are placed at $x=60a \pm d$. Note that the superconducting gap is at $E=\pm\Delta = 0.2t$.}
    \label{fig:LDOS_spatial}
\end{figure*}

\begin{figure*}[t]
    \centering
    \includegraphics[]{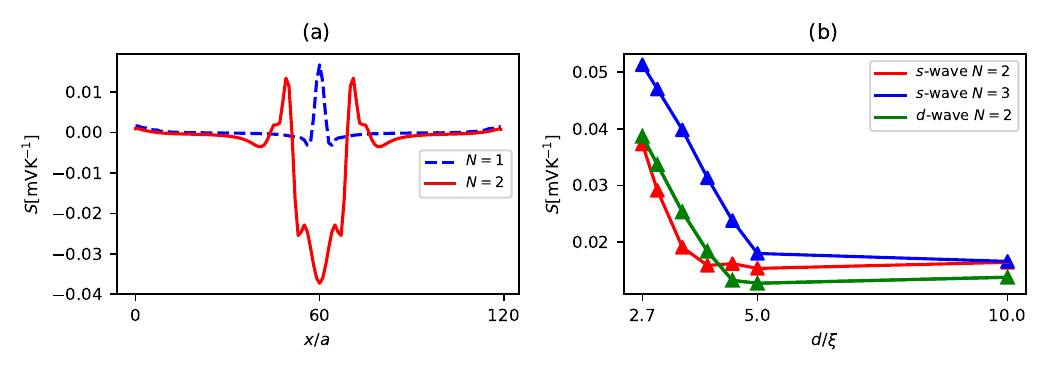}
    \caption{\textbf{(a)}: One-dimensional plot of the spatially resolved Seebeck coefficient along the intervortex line in the two-vortex case $N=2$, corresponding to the upper left plot of Fig. \ref{fig:OP_Seebeck_2D} (b). For comparison, we have included calculations for a superconductor with a single vortex ($N=1$) located at the center of the system. \textbf{(b)}: Maximum Seebeck coefficient attained in the superconductor as a function of the vortex separation distance. Here $N$ denotes the number of vortices present in the superconductor.}
    \label{fig:Seebeck_1D_Smax_vs_d}
\end{figure*}

The starting point is an extended Hubbard model with an attractive on-site ($s$-wave) or nearest-neighbor ($d$-wave) interaction between the electrons. Following a mean-field approximation, the Hamiltonian describing the superconductor is given by
\begin{equation}
\begin{split}
    H = \sum_{\langle i, j\rangle, \alpha}t_{ij} c_{i\alpha}^\dagger c_{j\alpha}^{\vphantom{\dagger}}  - \sum_{i\alpha} \mu_i c_{i \alpha}^\dagger c_{i\alpha}^{\vphantom{\dagger}} + H_s + H_d.
    \label{eq:hamiltonian}
\end{split}
\end{equation}
Here, the operators $c_{i\alpha}^\dagger$ ($c_{i\alpha}^{\vphantom{\dagger}}$) create (annihilate) electrons at lattice site $i$ with spin $\alpha$. The notation $\langle i, j \rangle$ denotes a sum over nearest-neighbors pairs. To model electric impurities, we make the chemical potential site-dependent, $\mu \to \mu_i = \mu - V_i^{\text{imp}}$, where $V_i^{\text{imp}}$ is the impurity potential. The last two terms are given by
\begin{align}
    H_s &= -U\sum_i (F_i^{\vphantom{\dag}} c_{i\uparrow}^\dagger c_{i\downarrow}^\dagger + \text{h.c.}),\\
    H_d &= -V\sum_{\langle i, j\rangle} (F_{ij}^{\vphantom{\dag}} c_{i\uparrow}^\dagger c_{j\downarrow}^\dagger + \text{h.c.}),
\end{align}
and give rise to $s$-wave and $d$-wave superconductivity, respectively. The superconducting pairing amplitudes are defined as $F_i=\langle c_{i\downarrow} c_{i\uparrow}\rangle $ for on-site pairing and $F_{ij} = \langle c_{j\downarrow} c_{i\uparrow} \rangle$ for nearest-neighbor pairing. The coupling parameters $U,V>0$ denote the strength of the attractive interaction between electrons causing formation of Cooper pairs. In our calculations we have set $U\neq0$ and $V=0$ to model $s$-wave superconductivity, and vice-versa to model $d$-wave superconductivity.

While electric impurities are accounted for in the model by the site-dependent chemical potential, vortices are induced by subjecting the system to an external magnetic field. Such a field is incorporated in the model by introducing a vector potential $\vec{A}(\vecr)$ in the Hamiltonian in (\ref{eq:hamiltonian}). This can be achieved in a simple manner by modifying the hopping parameter $t_{ij}\to t_{ij} e^{i\phi_{ij}}$ where $\phi_{ij} = -(e/\hbar) \int_{\vecR_j}^{\vecR_i} d\vecr \cdot \vec{A}(\vecr)$, a method known as Peierls substitution \cite{zhu_bogoliubov-gennes_2016}.  

The Hamiltonian is diagonalized by a Bogoliubov transformation, resulting in the following eigenvalue problem 
\begin{equation}
    E_n \begin{pmatrix}
        u_{ni}\\
        v_{ni}
    \end{pmatrix} = \sum_j\begin{pmatrix}
        t_{ij} - \mu_i \delta_{ij} & -\Delta_{ij}\\
        -\Delta_{ij}^* & -t_{ij}^* +\mu_i\delta_{ij}
    \end{pmatrix}
    \begin{pmatrix}
        u_{nj}\\
        v_{nj}
    \end{pmatrix}
    ,
\end{equation}
where we defined $\Delta_{ij}=UF_i\delta_{ij} + VF_{ij}$. This system of equations, the Bogoliubov-de Gennes (BdG) equations, is solved numerically and the pairing amplitudes are either determined self-consistently or pre-set to match known order-parameter (OP) profiles to speed up computation time for large systems. The local $s$-wave and $d$-wave OPs are defined from the pairing amplitudes respectively as
\begin{align}
    \Delta^s_i &= UF_i, \\
    \Delta^d_i &=  \frac{V}{4} (F_{i, i+\hat{\vec{x}}} + F_{i, i-\hat{\vec{x}}} - F_{i, i+\hat{\vec{y}}} - F_{i, i-\hat{\vec{y}}} ).
\end{align}
For the vortex calculations we have chosen to pre-set the OP, while for the impurity calculations it is determined self-consistently. The number of vortices can for instance be tuned by adjusting the applied magnetic field. However, a lower limit on the vortex separation distance between vortex cores exist. It has been predicted that for external fields close to the upper critical field, the corresponding minimum separation distance is approximately $2.7\xi$ \cite{poole_superconductivity_1995}.

The solutions to the BdG equations are then used to compute the LDOS according to the following formula:
\begin{equation}
    \rho(\vecr_i, E) = \sum_n \left[|u_{ni}|^2\delta(E-E_n) + |v_{ni}|^2\delta(E+E_n)\right],
\end{equation}
where the sum runs over all eigenvalues. In the numerical calculations we will approximate the Dirac-delta functions using normalized Gaussians of width $\sigma/t=0.01$. Finally, the result is used to calculate the Onsager elements using Eqs. (\ref{eq:L11}) and (\ref{eq:L12}). \\

The microscopic length scales accessible in real-space lattice Bogoliubov-de Gennes calculations place the present model in the quantum-limit regime where the spacing between the vortex-bound states is not negligible. Resolving several vortices within a finite lattice requires a relatively short coherence length due to computational restrictions. This results in values of $k_F\xi$ that are smaller than those typical of conventional weak-coupling metallic superconductors. This regime is nevertheless physically relevant to superconductors with a short coherence length and small Fermi energy, in which the spacing between the discrete vortex bound-states becomes sufficiently large for individual levels and their belonging particle-hole asymmetric spectral weights to be clearly experimentally resolved. Such discrete vortex-core levels have been observed by scanning tunneling spectroscopy in FeTe$_{0.55}$Se$_{0.45}$ and attributed to its small Fermi energy \cite{chen_natcom_18}. Particle-hole asymmetric vortex-core spectra associated with the quantum regime have also been reported in Ba$_{0.6}$K$_{0.4}$Fe$_2$As$_2$ and YNi$_2$B$_2$C \cite{shan_natphys_11, nishimori_jpsj_04}.

The present calculations should therefore be viewed as describing quantum-limit superconductors rather than as a rescaled quantitative model of a conventional large-$k_F\xi$ metal. Within this regime, the ratio of system size and vortex separation, $L/\xi$ and $d/\xi$, can still be chosen in experimentally standard ranges. This enables us to study how the overlap between spatially extended vortex-bound states evolves with vortex proximity. Our results therefore provide experimentally relevant predictions for quantum-limit superconductors and identify a qualitative interference mechanism that may remain operative at larger values of $k_F\xi$, although its magnitude and detailed spectral signatures are expected to become material dependent. In our computations, the selected system parameters correspond to a value of $k_F \xi = 12$. Decreasing it further substantially increases the magnitude of the Seebeck coefficient. To compute $k_F$ we assume the Fermi surface to be approximately circular, which is accurate for low values of the chemical potential $\mu$ in the tight-binding dispersion.

\section{Results: Vortex-bound states}
The OP for a superconductor with several vortices is chosen according to the following formula \cite{fu_iop_2025}
\begin{equation}
    \Delta(\vecr_i)=\Delta_0 \prod_n \tanh  \frac{|\vecr_i - \vecR_n|}{\xi_0} e^{i\theta_{in}},
\end{equation}
where $\vecR_n$ is the position of vortex $n$, and $\theta_{in}$ is the polar coordinate of $\vecr_i$ relative to $\vecR_n$, see Figure \ref{fig:OP_Seebeck_2D} (a). For the calculations in this section we have set $\mu/t=-1$ and $\Delta_0/t=0.2$.

\begin{figure*}[t]
    \centering
    \includegraphics[]{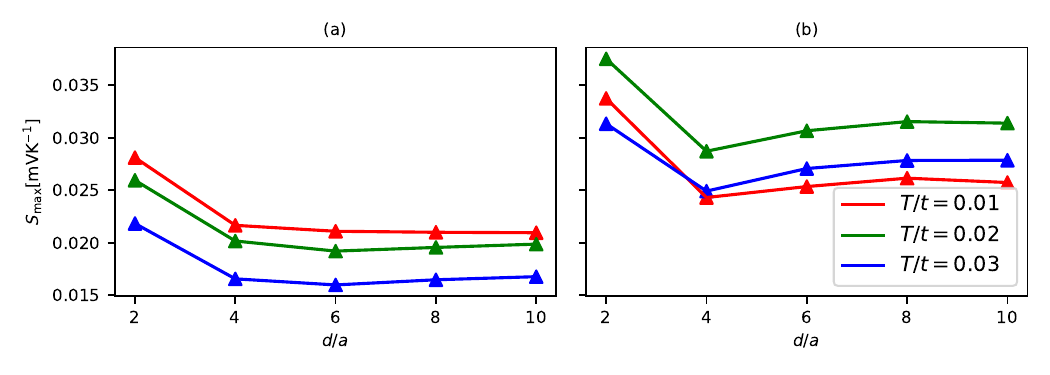}
    \caption{Maximum Seebeck coefficient in the (a) $s$-wave and (b) $d$-wave superconductor as a function of impurity separation distance for different temperatures.}
    \label{fig:Smax_vs_impsep}
\end{figure*}

As shown in Figure \ref{fig:OP_Seebeck_2D} (b), the calculations predict a large increase in the magnitude of the Seebeck coefficient in the intervortex region on the superconductor as the vortices are brought together in close proximity. The effect is clearly seen to be heavily dependent upon the vortex separation distance, with the upper right plot ($d/\xi=10$) indicating significantly smaller peak value centered around the individual vortices. For a small separation distance specifically, we observe about a four-fold increase in the Seebeck coefficient compared to the single-vortex case, see Figure \ref{fig:Seebeck_1D_Smax_vs_d} (a). Additionally, our calculations indicate that the effect is stronger in the three-vortex case. Figure \ref{fig:OP_Seebeck_2D} (b) indicates that the regions of the largest thermoelectric response is found along the lines joining the vortices. Figure \ref{fig:Seebeck_1D_Smax_vs_d} (b) explicitly shows the strong dependence on the vortex separation distance. Here we also see that the same effect appears in the $d$-wave superconductor. Compared to the $s$-wave superconductor, it performs quite similarly for small separation distances, while dropping off to a lower value as the vortices are moved further apart. 

As discussed above, the physical understanding behind the large thermoelectric response in the single-vortex case is that it occurs due to a large particle-hole asymmetry associated with the vortex-bound CdGM-states in the sub-gap region. However, when several vortices are brought together in close proximity, the CdGM-state wave functions overlap and hybridize to form new states. For two vortices, this is analogous to the overlap of two atomic orbitals, forming hybridized bonding and anti-bonding states. To visualize this, we present in Figure \ref{fig:LDOS_spatial} numerical calculations of the LDOS with small, medium and large vortex separations. Notice that as the separation distance is decreased, additional sub-gap states localized in the intervortex region appear. It is also apparent that the hybridized states break particle-hole symmetry, which we ultimately identify as the origin of the enhanced thermoelectric response. 

\begin{figure}[h]
    \centering
    \includegraphics[width=\linewidth]{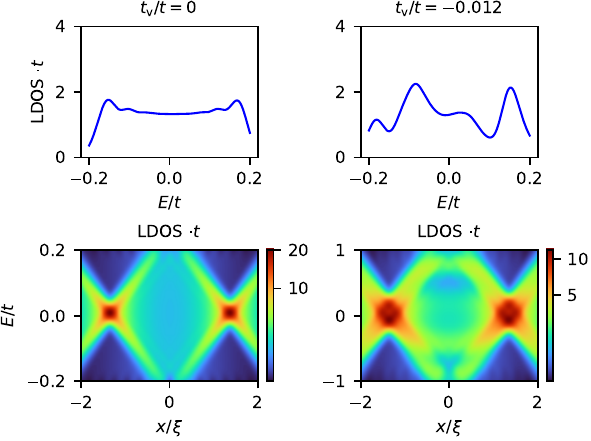}
    \caption{LDOS computed from the semi-analytical model. The lower plots are qualitatively similar to the corresponding numerically obtained results in Fig. \ref{fig:LDOS_spatial}. The upper plots show the LDOS specifically at the midway point between the vortices, showing enhanced particle-hole asymmetry as the interaction between the two vortices is turned on ($t_\text{V}\neq 0$.)} 
    \label{fig:analytical_ldos}
\end{figure}

To verify our interpretation of the numerical results, we investigate a semi-analytical model describing explicitly hybridization of vortex-bound states. For this purpose, consider as a basis the set of isolated vortex-bound states of energy $\epsilon_n$ localized around vortex $i$, $\{ |\psi_{in} \rangle \}$. We can then write down the general Hamiltonian in terms of these basis states \cite{fu_iop_2025}
\begin{equation}
    H = \sum_{in} \epsilon_n |\psi_{in} \rangle \langle \psi_{in} | + \sum_{i\neq j, nn'}t_{ij, nn'} | \psi_{in} \rangle \langle\psi_{jn'}|,
\end{equation}
where the parameters $t_{ij, nn'}$ encode interaction effects stemming from non-zero overlap between vortex-bound states. To keep things simple and analytically tractable while qualitatively capturing the essential physics of the problem, we limit ourselves to two-site vortex lattice. Moreover, we only consider hybridization between states for which $n=n'$, hence setting $t_{ij,n\neq n'}=0$. In this subspace, the Hamiltonian is reduced to
\begin{equation}
    H = \sum_n\begin{pmatrix}
        \epsilon_n & -t_\text{V} \\
        -t_\text{V} & \epsilon_n
    \end{pmatrix}
    ,
\end{equation}
with the overlap parameter $t_\text{V}>0$ chosen to fit numerical results. The sign-choice of $t_\text{V}$ follows from requiring the ground state wave function to have no nodes. The eigenstates are
\begin{equation}
    \tilde{u}_n^{\pm} = \frac{1}{\sqrt{2}} 
    \begin{pmatrix}
         1\\\pm 1 
    \end{pmatrix}
\end{equation}
with corresponding eigenvalues $E_n^{\pm}= \epsilon_n \mp t_\text{V}$. The LDOS can then be calculated from the CdGM wave functions as
\begin{equation}
    \rho(\vecr, E) = \sum_{n, \sigma=\pm 1} |u_n(\vecr-\vecR_1) + \sigma u_n(\vecr - \vecR_2)|^2 \delta(E - \epsilon_n  +\sigma t_\text{V}),
\end{equation}
where $\vecR_{1,2}$ denote the positions of the vortices. We use the analytical CdGM wave functions valid for the low-lying excitations, which are given by \cite{caroli_bound_1964}
\begin{equation}
    u_n(r, \theta) = A e^{-K(r)} J_{n-1/2}(k_F r) e^{i(n-1/2)\theta}.
\end{equation}
Since we are looking for qualitative agreement with the numerical results, we set the normalization constant $A=1$ for simplicity. We plot the resulting LDOS in the two-site vortex lattice in Fig. \ref{fig:analytical_ldos}.
Despite its simplicity, the model qualitatively reproduces the main features of the full numerical BdG results in Fig. \ref{fig:LDOS_spatial}. In particular, finite overlap splits the isolated-vortex levels into bonding and antibonding combinations and generates additional spectral weight in the region between the vortices. The interference between the two spatially displaced wave functions also produces a spatially dependent redistribution of spectral weight, demonstrating that the multi-vortex spectrum cannot be obtained by simply superposing the LDOS of two isolated vortices. The quantitative energies and intensities differ from the numerical results because the analytical model retains only the low-energy CdGM wave functions and uses a single phenomenological overlap parameter. Nevertheless, the qualitative agreement between Figs. \ref{fig:LDOS_spatial} and \ref{fig:analytical_ldos} supports our interpretation that the separation-dependent reconstruction of the LDOS originates from hybridization and interference between vortex-bound states.

\section{Results: Impurity-bound states} 

\begin{figure*}[t]
    \centering
    \includegraphics[]{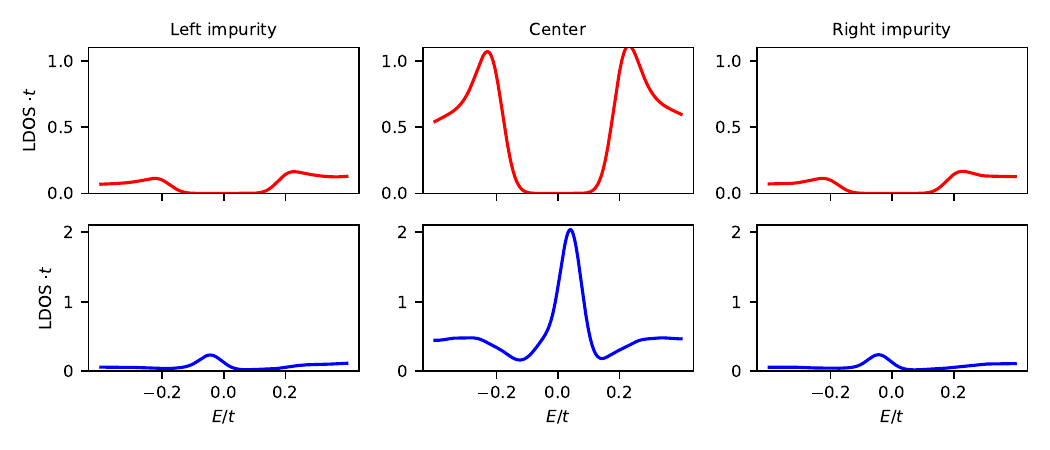}
    \caption{LDOS for the two-impurity configuration with separation $d/a=2$ in a $s$-wave (upper panel) and $d$-wave (lower panel) superconductor.}
    \label{fig:LDOS_impurity_hybridization}
\end{figure*}

We now move on to present self-consistent calculations for $s$- and $d$-wave superconductors subject to electric impurities. For the calculations in this section we have set $U/t = 1.5$, $V/t = 0.4$, $\mu=0$ and the impurity strength $V_{\text{imp}}/t=3$. Limiting ourselves to the case of two impurities placed at a separation $d$ from each other, we plot in Figure \ref{fig:Smax_vs_impsep} the maximum Seebeck coefficient attained in the superconductor as a function of this separation distance. In both the $s$- and $d$-wave cases we find that as the proximity of the impurities is altered, the maximum Seebeck coefficient also changes. When the impurities are placed two lattice sites apart, we see a maximum relative increase of about 30\% compared to the single-impurity setup in both cases. We note that this is significantly smaller than the increase we saw in the vortex-case. The calculations also indicate a separation interval in which the thermoelectric response is reduced. Somewhat surprisingly, our calculations predict a non-zero and indeed sizable thermoelectric response in the $s$-wave superconductor, despite the energy of the induced bound state being at the gap edge as previously discussed. 

While the calculations indicate that the thermoelectric response is altered, even enhanced for small enough separation distances, we do not consider this conclusive proof that hybridization effects are the origin of this. To investigate this further, one would have to look closer at how the electronic structure is modified by the presence of several impurities. To this end we have included in Figure \ref{fig:LDOS_impurity_hybridization} plots of the LDOS in both the $s$- and $d$-wave cases. In the $s$-wave case, there are seemingly no hybridization effects enhancing particle-hole asymmetry in the LDOS. This is reasonable because the single-impurity bound state does not extend beyond the impurity site \cite{balatsky_impurity-induced_2006}. We suspect that the reason we nevertheless see an increase in the Seebeck coefficient is not directly due to hybridization effects, but rather that the individual effects of the impurities combined boost thermoelectricity. Recall that the normal metal is modeled as an STM tip which is still sensitive to surrounding sites, and therefore it is plausible that we see an increase in the Seebeck coefficient when the impurity sites are close, even without hybridization effects. In the $d$-wave superconductor the situation is different. For the case when the impurities are separated by two lattice constants we find a sharp peak in the LDOS at the midway point, indicating a hybridization effect between the bound states. Note that in this case the midway point corresponds to a nearest-neighbor site to both impurity sites. As the bound state belonging to a single impurity extends to the nearest-neighbor sites \cite{balatsky_impurity-induced_2006}, it is therefore not surprising that when the impurities are separated by only two lattice constants we see a hybridization effect.

\section{Concluding remarks} 

We have investigated how the local thermoelectric response of superconductors is modified by the proximity of multiple vortices or nonmagnetic impurities. For vortices, bringing the cores closer together produces a pronounced enhancement of the maximum Seebeck coefficient. The effect extends beyond the immediate core regions and is accompanied by a reconstruction of the subgap spectrum in the intervortex region. A minimal two-vortex model qualitatively reproduces the level splitting and spatial redistribution of spectral weight obtained from the lattice BdG calculations, supporting hybridization and interference between vortex-bound states as the origin of the enhanced particle-hole asymmetry. For nonmagnetic impurities, both the $s$- and $d$-wave superconductors exhibit a nonmonotonic dependence of the thermoelectric response on impurity separation, with a maximum enhancement of approximately $30\%$ relative to the single-impurity case for the parameters considered. The associated spectral reconstruction is nevertheless qualitatively different. In the $s$-wave case it remains strongly localized to the impurity sites, whereas in the $d$-wave case appreciable bound-state weight extends into the interimpurity region. The latter behavior is consistent with overlap between the impurity-induced states, while the microscopic origin of the $s$-wave enhancement remains less conclusive. These results demonstrate that defect separation provides a means of controlling local thermoelectricity through the spatial structure and interaction of defect-induced quasiparticle states. The effect should be particularly relevant in quantum-limit superconductors, where discrete vortex-bound states and their particle-hole-asymmetric spectral weights are experimentally resolvable.

\acknowledgments 
This work was supported by the Research Council
of Norway through Grant No. 353894 and its Centres of
Excellence funding scheme Grant No. 262633 “QuSpin.”
Support from Sigma2 - the National Infrastructure for High
Performance Computing and Data Storage in Norway, project
NN9577K, is acknowledged.\\

\textit{Disclosure of Generative AI use:} During the preparation
of this manuscript, the authors used generative AI tools. Microsoft Copilot GPT 5.6 was used as interactive brainstorming aids, for style editing of portions of the text and for generating the schematic model figure 1. The same tool was also used in the development of numerical code, primarily providing assistance with debugging, optimization, and code structure. All AI-generated
outputs were thoroughly reviewed, verified, and edited by
the authors, who maintain full responsibility for the scientific
integrity, accuracy, and conclusions of this work.


\bibliography{references}

@article{nishimori_jpsj_04,
  author  = {Nishimori, Hitoshi and Uchiyama, Kazuharu and Kaneko, Shin-ichi
             and Tokura, Akio and Takeya, Hiroyuki and Hirata, Kazuto
             and Nishida, Nobuhiko},
  title   = {First Observation of the Fourfold-Symmetric and Quantum-Regime
             Vortex Core in {YNi}$_2${B}$_2${C} by Scanning Tunneling
             Microscopy and Spectroscopy},
  journal = {Journal of the Physical Society of Japan},
  volume  = {73},
  number  = {12},
  pages   = {3247--3250},
  year    = {2004},
  doi     = {10.1143/JPSJ.73.3247}
}

@article{shan_natphys_11,
  author  = {Shan, Lei and Wang, Yong-Lei and Shen, Bing and Zeng, Bin
             and Huang, Yan and Li, Ang and Wang, Da and Yang, Huan
             and Ren, Cong and Wang, Qiang-Hua and Pan, Shuheng
             and Wen, Hai-Hu},
  title   = {Observation of ordered vortices with {Andreev} bound states
             in {Ba}$_{0.6}${K}$_{0.4}${Fe}$_2${As}$_2$},
  journal = {Nature Physics},
  volume  = {7},
  pages   = {325--331},
  year    = {2011},
  doi     = {10.1038/nphys1908}
}

@article{chen_natcom_18,
author = {Chen, Mingyang and Chen, Xiaoyu and Yang, Huan and Du, Zengyi
and Zhu, Xiyu and Wang, Enyu and Wen, Hai-Hu},
title = {Discrete energy levels of {Caroli--de Gennes--Matricon} states
in the quantum limit in {FeTe}$_{0.55}${Se}$_{0.45}$},
journal = {Nature Communications},
volume = {9},
pages = {970},
year = {2018},
doi = {10.1038/s41467-018-03404-8}
}

@book{tinkham_book,
author = {Tinkham, M.},
publisher = {Dover Books on Physics Series, Dover, New York},
title = {{Introduction to Superconductivity}},
year = {2004}
}

@article{heikkila_pss_19, 
  author={Heikkilä, T. and Silaev, M. and Virtanen, M. and Bergeret, F. S.},
  title={Thermal, electric and spin transport in superconductor/ferromagnetic-insulator structures}, 
  journal={Prog. Surf. Sci.}, 
  volume={94},
  pages={100540},
  year={2019},
url={https://www.sciencedirect.com/science/article/abs/pii/S0079681619300115}
}

@article{marchegiani2020nonlinear,
  title={Nonlinear thermoelectricity with electron-hole symmetric systems},
  author={Marchegiani, G and Braggio, A and Giazotto, F},
  journal={Physical review letters},
  volume={124},
  number={10},
  pages={106801},
  year={2020},
  publisher={APS},
url={https://journals.aps.org/prl/abstract/10.1103/PhysRevLett.124.106801}
}

@article{singh_prl_2024,
  title = {Giant Thermoelectric Response of Fluxons in Superconductors},
  author = {Singh, Alok Nath and Bhandari, Bibek and Braggio, Alessandro and Giazotto, Francesco and Jordan, Andrew N.},
  journal = {Phys. Rev. Lett.},
  volume = {133},
  issue = {25},
  pages = {256002},
  numpages = {7},
  year = {2024},
  month = {Dec},
  publisher = {American Physical Society},
  doi = {10.1103/PhysRevLett.133.256002},
  url = {https://link.aps.org/doi/10.1103/PhysRevLett.133.256002}
}

@article{fu_iop_2025,
  title = {Quantum interference among vortex bound states in superconductors},
  author = {Fu, Yi Mingh and Wang, Da and Wang, Qiang Hua},
  journal = {New J. Phys.},
  volume = {27},
  issue = {7},
  pages = {073502},
  year = {2025},
  month = {Jul},
  doi = {10.1088/1367-2630/adea18},
  url = {https://iopscience.iop.org/article/10.1088/1367-2630/adea18}
}

@article{abrikosov_magnetic_1957,
	title = {The magnetic properties of superconducting alloys},
	volume = {2},
	copyright = {https://www.elsevier.com/tdm/userlicense/1.0/},
	issn = {00223697},
	url = {https://linkinghub.elsevier.com/retrieve/pii/0022369757900835},
	doi = {10.1016/0022-3697(57)90083-5},
	number = {3},
	urldate = {2026-04-28},
	journal = {Journal of Physics and Chemistry of Solids},
	author = {Abrikosov, A.A.},
	month = jan,
	year = {1957},
	pages = {199--208},
}

@article{caroli_bound_1964,
    title = {Bound {Fermion} states on a vortex line in a type {II} superconductor},
    volume = {9},
    copyright = {https://www.elsevier.com/tdm/userlicense/1.0/},
    issn = {00319163},
    url = {https://linkinghub.elsevier.com/retrieve/pii/0031916364903750},
    doi = {10.1016/0031-9163(64)90375-0},
    number = {4},
    urldate = {2026-02-04},
    journal = {Physics Letters},
    author = {Caroli, C. and De Gennes, P.G. and Matricon, J.},
    month = may,
    year = {1964},
    pages = {307--309},
}

@article{qin_solid-state_2022,
    title = {Solid-state cooling: thermoelectrics},
    volume = {15},
    issn = {1754-5692, 1754-5706},
    shorttitle = {Solid-state cooling},
    url = {https://xlink.rsc.org/?DOI=D2EE02408J},
    doi = {10.1039/D2EE02408J},
    number = {11},
    urldate = {2026-05-11},
    journal = {Energy \& Environmental Science},
    author = {Qin, Yongxin and Qin, Bingchao and Wang, Dongyang and Chang, Cheng and Zhao, Li-Dong},
    year = {2022},
    pages = {4527--4541},
}

@book{zhu_bogoliubov-gennes_2016,
    series = {Lecture {Notes} in {Physics}},
    title = {Bogoliubov-de {Gennes} {Method} and {Its} {Applications}},
    volume = {924},
    copyright = {http://www.springer.com/tdm},
    isbn = {978-3-319-31312-2 978-3-319-31314-6},
    url = {http://link.springer.com/10.1007/978-3-319-31314-6},
    doi = {10.1007/978-3-319-31314-6},
    urldate = {2025-10-15},
    publisher = {Springer International Publishing},
    author = {Zhu, Jian-Xin},
    year = {2016},
}

@book{goldsmid_introduction_2016,
    address = {Berlin, Heidelberg},
    series = {Springer {Series} in {Materials} {Science}},
    title = {Introduction to {Thermoelectricity}},
    volume = {121},
    copyright = {https://www.springer.com/tdm},
    isbn = {978-3-662-49255-0 978-3-662-49256-7},
    url = {https://link.springer.com/10.1007/978-3-662-49256-7},
    doi = {10.1007/978-3-662-49256-7},
    urldate = {2026-05-08},
    publisher = {Springer Berlin Heidelberg},
    author = {Goldsmid, H. Julian},
    year = {2016},
}

@article{benenti_fundamental_2017,
    title = {Fundamental aspects of steady-state conversion of heat to work at the nanoscale},
    volume = {694},
    issn = {03701573},
    url = {https://linkinghub.elsevier.com/retrieve/pii/S0370157317301540},
    doi = {10.1016/j.physrep.2017.05.008},
    urldate = {2026-08-28},
    journal = {Physics Reports},
    author = {Benenti, Giuliano and Casati, Giulio and Saito, Keiji and Whitney, Robert S.},
    month = jun,
    year = {2017},
    pages = {1--124},
}

@article{machida_bound_1972,
    title = {Bound {States} {Due} to {Resonance} {Scattering} in {Superconductor}},
    volume = {47},
    issn = {0033-068X},
    url = {https://academic.oup.com/ptp/article-lookup/doi/10.1143/PTP.47.1817},
    doi = {10.1143/PTP.47.1817},
    number = {6},
    urldate = {2026-03-13},
    journal = {Progress of Theoretical Physics},
    author = {Machida, Kazushige and Shibata, Fumiaki},
    month = jun,
    year = {1972},
    pages = {1817--1823},
}

@article{shiba_hartree-fock_1973,
    title = {A {Hartree}-{Fock} {Theory} of {Transition}-{Metal} {Impurities} in a {Superconductor}},
    volume = {50},
    issn = {0033-068X},
    url = {https://academic.oup.com/ptp/article-lookup/doi/10.1143/PTP.50.50},
    doi = {10.1143/PTP.50.50},
    number = {1},
    urldate = {2026-03-13},
    journal = {Progress of Theoretical Physics},
    publisher = {Progress of Theoretical Physics},
    author = {Shiba, Hiroyuki},
    month = jul,
    year = {1973},
    pages = {50--73},
}

@article{balatsky_impurity-induced_2006,
    title = {Impurity-induced states in conventional and unconventional superconductors},
    volume = {78},
    copyright = {http://link.aps.org/licenses/aps-default-license},
    issn = {0034-6861, 1539-0756},
    url = {https://link.aps.org/doi/10.1103/RevModPhys.78.373},
    doi = {10.1103/RevModPhys.78.373},
    number = {2},
    urldate = {2025-10-14},
    journal = {Reviews of Modern Physics},
    author = {Balatsky, A. V. and Vekhter, I. and Zhu, Jian-Xin},
    month = may,
    year = {2006},
    pages = {373--433},
}

@article{lofwander_large_2004,
    title = {Large thermoelectric effects in unconventional superconductors},
    volume = {70},
    copyright = {http://link.aps.org/licenses/aps-default-license},
    issn = {1098-0121, 1550-235X},
    url = {https://link.aps.org/doi/10.1103/PhysRevB.70.024515},
    doi = {10.1103/PhysRevB.70.024515},
    number = {2},
    urldate = {2025-10-14},
    journal = {Physical Review B},
    author = {Löfwander, T. and Fogelström, M.},
    month = jul,
    year = {2004},
    pages = {024515},
}

@book{poole_superconductivity_1995,
    address = {San Diego},
    title = {Superconductivity},
    isbn = {978-0-12-561455-9 978-0-12-561456-6},
    publisher = {Academic Press},
    author = {Poole, Charles P. and Farach, Horacio A. and Creswick, Richard J.},
    year = {1995},
}

\end{document}